\documentclass[superscriptaddress,twocolumn,showpacs,preprintnumbers,amsmath,amssymb,floatfix,nofootinbib]{revtex4-2}

\usepackage{color}   
\usepackage{graphicx}
\usepackage{dcolumn}

\begin{document}

\def\bb    #1{\hbox{\boldmath${#1}$}}

\title{Relaxation and correlation times of nonequilibrium multiparticle systems}

\author{Maciej Rybczy\'nski}
\email{maciej.rybczynski@ujk.edu.pl}
\affiliation{Institute of Physics, Jan Kochanowski University, 25-406 Kielce, Poland}
\author{Grzegorz Wilk}
\email{grzegorz.wilk@ncbj.gov.pl}
\affiliation{ National Centre for Nuclear Research, Warsaw 02-093, Poland}
\author{Zbigniew W\l odarczyk}
\email{zbigniew.wlodarczyk@ujk.edu.pl}
\affiliation{Institute of Physics, Jan Kochanowski University, 25-406 Kielce, Poland}

\begin{abstract}
Relaxation and correlation times are two parameters used frequently in approximate descriptions of the time development of hadronizing system from some initial state towards distributions observed experimentally. 
Chosen to reproduce the experimental results they represent, in a sense, the history of the hadronization process.
The analysis of their changes with energy is the subject of our work.
\end{abstract}


\maketitle
\section{Introduction}
\label{Introduction}

In many descriptions of multiparticle production processes we are interested in their temporal development from a certain initial state to the final state recorded in the experiment. In such descriptions, we usually deal with certain multi-stage processes, each with its own characteristic time scale. The concept of their hierarchy is one of the fundamental properties in statistical physics \cite{BK}. These  phenomena  can  be  understood correctly  only if the  dynamics of one-particle and two-particle properties characterized by, respectively, relaxation time $\tau_{rel}$ and correlation time $\tau_{cor}$, are known. In most situations first, in the initial stage, i.e., for $t < \tau_{cor}$ the correlations relax, it is followed in $\tau_{cor} < t < \tau_{rel}$ by the kinetic stage when the one-particle relax, and finally, for $t> \tau_{rel}$ system enters into stationary (hydrodynamic) stage. In this way the relaxation and correlation times (chosen to reproduce the experimental results for, respectively, transverse distributions of produced secondaries and multiplicity distributions) occur as two parameters which represent the dynamical history of the hadronization process. As such they must depend on energy and the form and details of this dependence is the subject of this work.

The evolution of the particle distribution can be studied through the Boltzmann transport equation (BTE),
\begin{equation}
\frac{df(r,p,t)}{dt} = \frac{\partial f}{\partial t} + \vec{u}\cdot\nabla_r f + \vec{F}\cdot \nabla_p f = C[f], \label{BE}
\end{equation}
where $f(r,p,t)$ is the distribution of particles which depends on position $r$, momentum $p$ and time $t$, $\vec{F}$ is the external force, $\vec{u}$ is the velocity and $C[f]$ is the collision term. Assuming in what follows homogeneity of the system ($\nabla_r f=0$) and absence of external forces ($\vec{F}=0$) Eq. (\ref{BE}) reduces to
\begin{equation}
\frac{df(r,p,t)}{dt} = \frac{\partial f}{\partial t} = C[f]. \label{RBE}
\end{equation}
In the relaxation time approximation (RTA) \cite{BGK,AW,Balescu,FR} the collision term is assumed to be equal to
\begin{equation}
C[f] = \frac{f_{eq} - f}{\tau_{rel}}, \label{RTA}
\end{equation}
where $f_{eq}$ is the local equilibrium distribution and $\tau_{rel}$ is the relaxation time, understood as the time taken by the non-equilibrium system to reach equilibrium. In this approximation BTE simplifies to
\begin{equation}
\frac{\partial f}{\partial t} = \frac{f_{eq} - f}{\tau_{rel}}. \label{SRBE}
\end{equation}
Solving this equation for the initial conditions such that at $t=0$ one has initial distribution, $f = f_{in}$, and at freeze-out time, $t=t_f$ one has final distribution, $f=f_{fin}$ (to be identified with the actually measured distribution) one gets that
\begin{equation}
f_{fin} = f_{eq} + \left(f_{in} - f_{eq}\right) \exp\left( - \frac{t_f}{\tau_{rel}}\right). \label{RTA-sol}
\end{equation}
The Boltzmann transport equation in the RTA approximation is a very popular approach recently used to analyze the various observables from nucleus-nucleus collisions measured in experiments at RHIC and LHC , cf., for example, \cite{TKTS,TBGKSC,BGSS,ZLD,YTTS,QCGZZ}\footnote{BTE in RTA approximation has been used to study the time evolution of temperature fluctuations in a non-equilibrated system \cite{BGSS}, elliptic flow \cite{YTTS} and also for study nuclear modification \cite{TKTS,TBGKSC,QCGZZ} factor at RHIC and LHC energies.}.

\vspace*{-0.2cm}

\section{Deducing energy dependence of $t_f/\tau_{rel}$ from data on $p_T$ distributions}
\label{Distribution}

 We shall analyse in this work transverse momentum distributions, $f\left(p_T\right)$, from proton-proton and proton-antiproton collisions in a wide range of energies. To start let us note that we need $f_{fin}$ which can be identified with the experimentally distribution. Therefore we need as good as possible formula fitting $p_T$ at all energies available. We argue that such a formula is the Tsallis power-law distribution \cite{WW1,WWCT,WW2}
\begin{equation}
f\left( p_T\right) = \frac{2-q}{T}\left[ 1 + (q-1)\frac{p_T}{T}\right]^{\frac{1}{1-q}} \label{TFit}
\end{equation}
characterized by the energy dependent Tsallis $q$ parameter and the temperature parameter $T$. Here $q \ge 1$, for $q \to 1$ Tsallis distribution becomes usual Boltzmann-Gibbs distribution,
\begin{equation}
f_{BG}\left(p_T\right) = \frac{1}{T}\exp\left( - \frac{p_T}{T}\right). \label{BG}
\end{equation}
As shown in \cite{WWCT} this formula nicely describes wide range of the measured transverse momenta ($0.1 < p_T < 100$ GeV) in which cross section spans a range of $\sim 14$ orders of magnitude. Parameter $q$ represents the degree of the non-extensivity or, in other words, the degree of deviation of the system from the thermalized or equilibrated system, which is usually described by the well known Boltzmann-Gibbs statistical mechanics. In our case, it is a limiting form of the considered system for $t_f \to \infty$. Therefore our $f_{eq}$ in Eq. (\ref{RTA-sol}) is assumed to have form of Eq. (\ref{BG}).

The above-mentioned features of the Tsallis distribution (see also \cite{WW2}) mean that also $f_{in}$ can be selected in this form, but with $q$ characteristic for hard scattering. Its value can be estimated by assuming the basic quark model as responsible for the initial state. In this case the high $p_T$ differential cross section can be inferred from the counting rules \cite{BF1,BF2,MMT} stating that for such processes the invariant cross section for the exclusive process at high $p_T$ behaves as the power law, with power index $\gamma=2\times[ (\rm{number~of~active~participants})-2]$. Assuming that the dominant processes of this type are $2\to 2$ processes (like $qq \to qq$) one gets that $d\sigma /dp_T \propto p_T^{-\gamma}$ with $\gamma = 4$, what translates to $q_{in} - 1 = 1/\gamma $.

\begin{figure}[h]
\begin{center}
\includegraphics[scale=0.43]{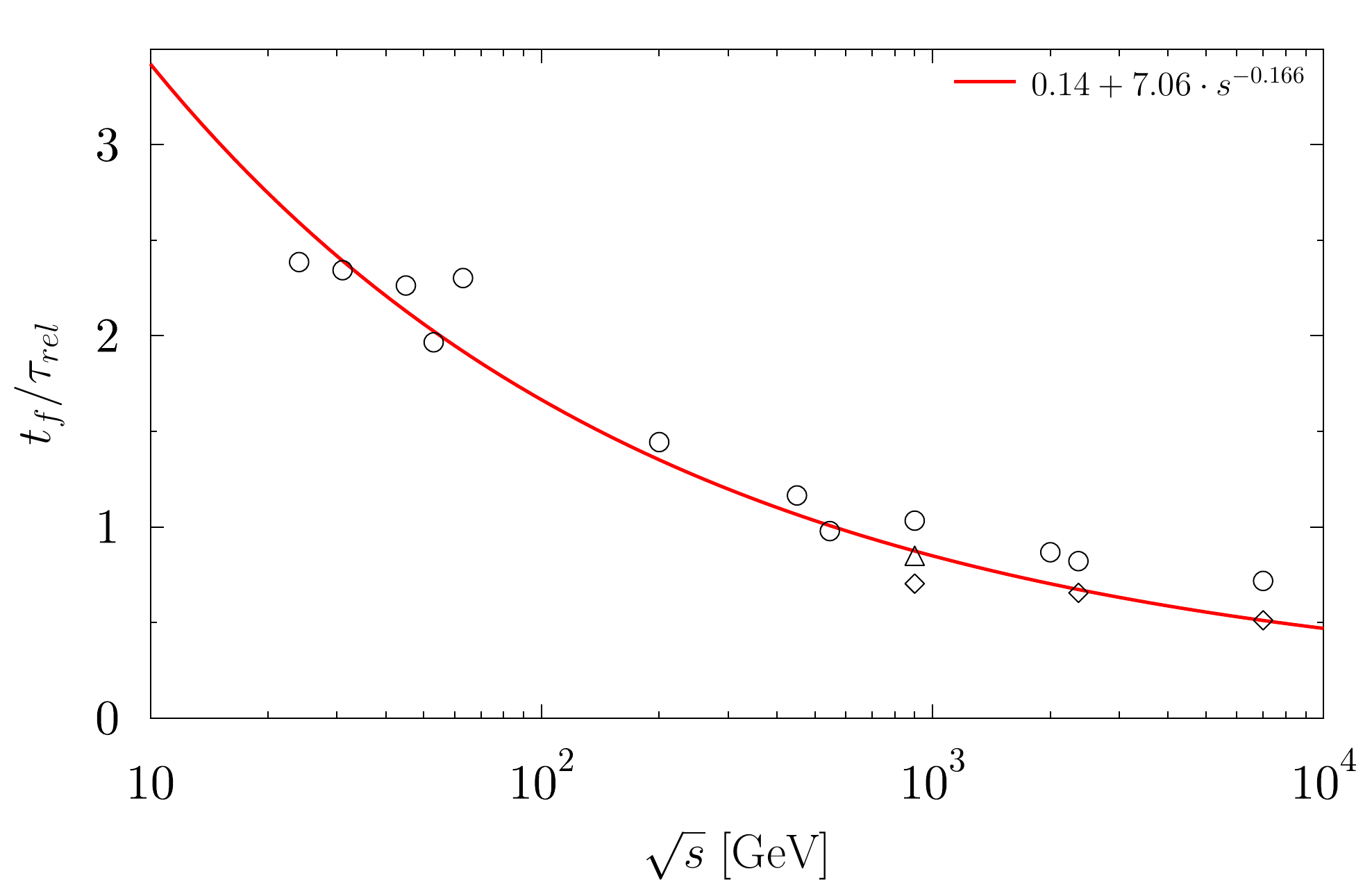}
\end{center}
\vspace{-5mm}
\caption{Energy dependence of $t_f/\tau_{rel}$ obtained from the experimental data using Eq. (\ref{Result}). Based on data from: \cite{WWCT} (triangle), \cite{Wybig} (circles) and \cite{RW} (diamonds).}
\label{F1}
\end{figure}

To find the dependence of the $t_f/\tau_{rel}$ ratio on energy we calculate the relation between temperatures deduced from different components of Eq. (\ref{RTA-sol}) using the fact that for Tsallis distribution
\begin{equation}
\langle p_T\rangle = \frac{T}{3-2q}. \label{qmeanpT}
\end{equation}
Using this in Eq. (\ref{RTA-sol}) one obtains that
\begin{equation}
\langle p_T\rangle (3 - 2q) = \langle p_T\rangle + \left[ \langle p_T\rangle \left(3-2q_{in}\right) - \langle p_T\rangle \right]\cdot \exp \left(-\frac{t_f}{\tau_{rel}}\right) \label{sumrule}
\end{equation}
and assuming that $\langle p_T\rangle = const$ during the time evolution one gets that
\begin{equation}
\frac{t_f}{\tau_{rel}} = \ln \left( \frac{q_{in} - 1}{q-1}\right).               \label{Result}
\end{equation}
Using for $q = q(s)$ values obtained from the experimental data on transverse momentum distributions for different energies \cite{Wybig,WWCT,RW} we obtain the ratio $t_f/\tau_{rel}$ as shown in Fig. \ref{F1}. Closing this Section, let us note that assuming that all distributions used here are Tsallis distributions, we are actually going beyond the RTA scheme (see Appendix for details).

\section{Deducing energy dependence of $t_f/\tau_{cor}$ from data on multiplicity}
\label{Multiplicity}

We will now move on to correlation time $\tau_{cor}$ which determines multiplicity distribution $P(N)$ \cite{Balescu}. Its scaled variance is given by the correlation function $\nu_2\left(t_1,t_2\right) = \nu_2\left(t=\left|t_1 - t_2\right|\right)$ by the relation \cite{Good}
\begin{equation}
\frac{{\rm Var}\left(N\right)}{\langle N\rangle} = 1 + \langle N\rangle \langle \nu_2\rangle, \label{VarN}
\end{equation}
where
\begin{equation}
\!\! \langle \nu_2\rangle = \int\!\!\int \nu_2\left( t_1,t_2\right)dt_1 dt_2 = \frac{2}{t^2_f}\int_0^{t_f} \left( t_f - t\right)\nu_2(t) dt.           \label{mu2}
\end{equation}
For the correlation function of the  the form
\begin{equation}
\nu_2(t) = \exp\left( - \frac{2 t}{\tau_{cor}}\right)  \label{taucor}
\end{equation}
one gets
\begin{equation}
\langle \nu_2\rangle = \left( \frac{\tau_{cor}}{t_f}\right)^2\left[ \exp\left( - \frac{t_f}{\tau_{cor}}\right) - 1 + 2 \frac{t_f}{\tau_{cor}} \right] \label{nu2}
\end{equation}
and the scaled variance is equal to \footnote{Notice that for multiplicity distribution expressed via Negative Binomial form is characterized by the parameter $\frac{1}{k} = \frac{{\rm Var}\left(N\right)}{\langle N\rangle} - \frac{1}{\langle N\rangle}  = \langle \nu_2\rangle$.}
\begin{equation}
\frac{{\rm Var}\left(N\right)}{\langle N\rangle} = 1 + \frac{\langle N\rangle}{2}\left( \frac{\tau_{cor}}{t_f}\right)^2\left[ \exp\left( - \frac{t_f}{\tau_{cor}}\right) - 1 + 2 \frac{t_f}{\tau_{cor}} \right]. \label{VarN-1}
\end{equation}
\begin{figure}[h]
\begin{center}
\includegraphics[scale=0.43]{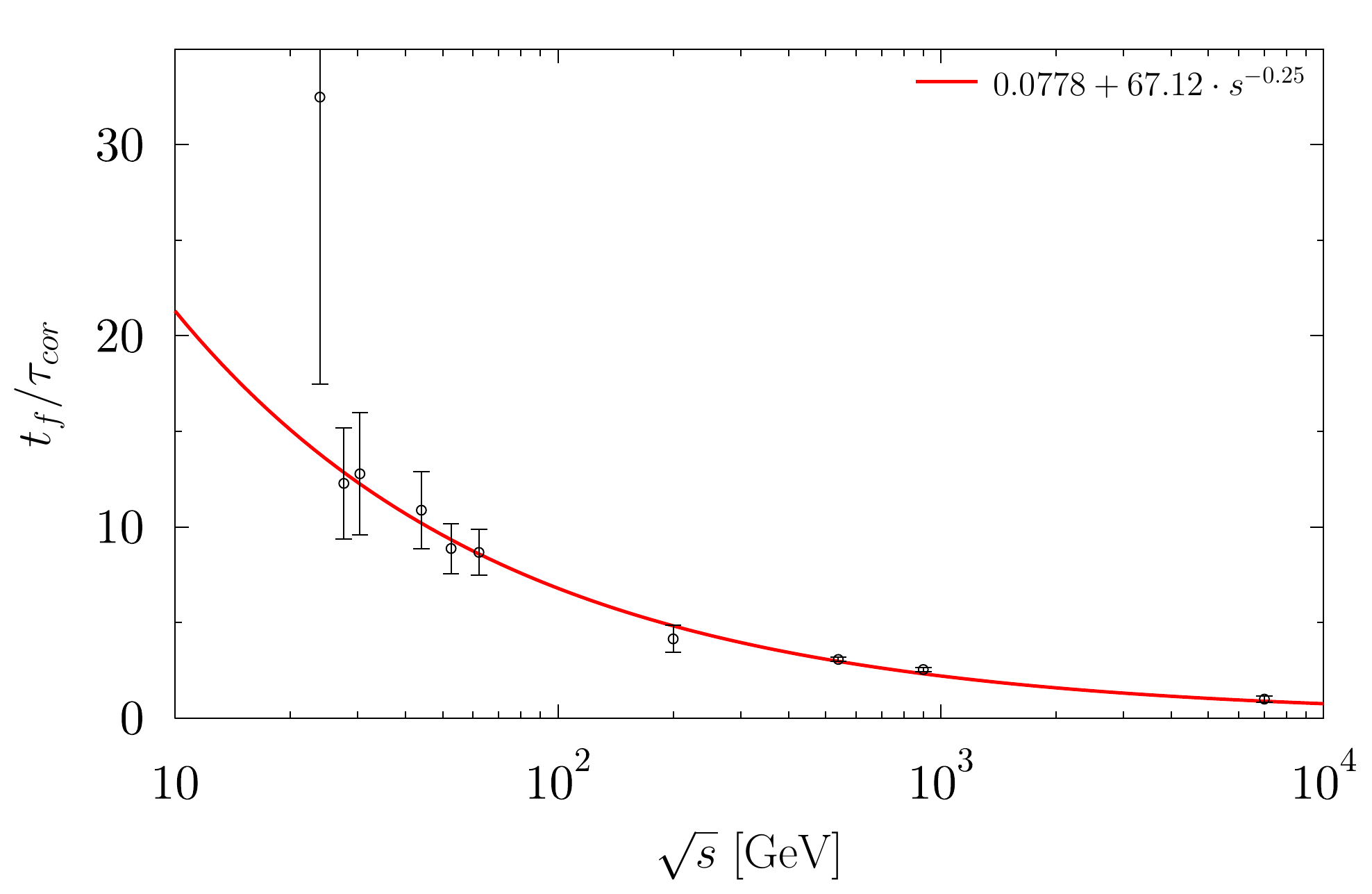}
\end{center}
\vspace{-5mm}
\caption{Energy dependence of $t_f/\tau_{cor}$ obtained from experimental data using Eq. (\ref{VarN-1}). }
\label{F2}
\end{figure}
\begin{figure}[h]
\begin{center}
\includegraphics[scale=0.43]{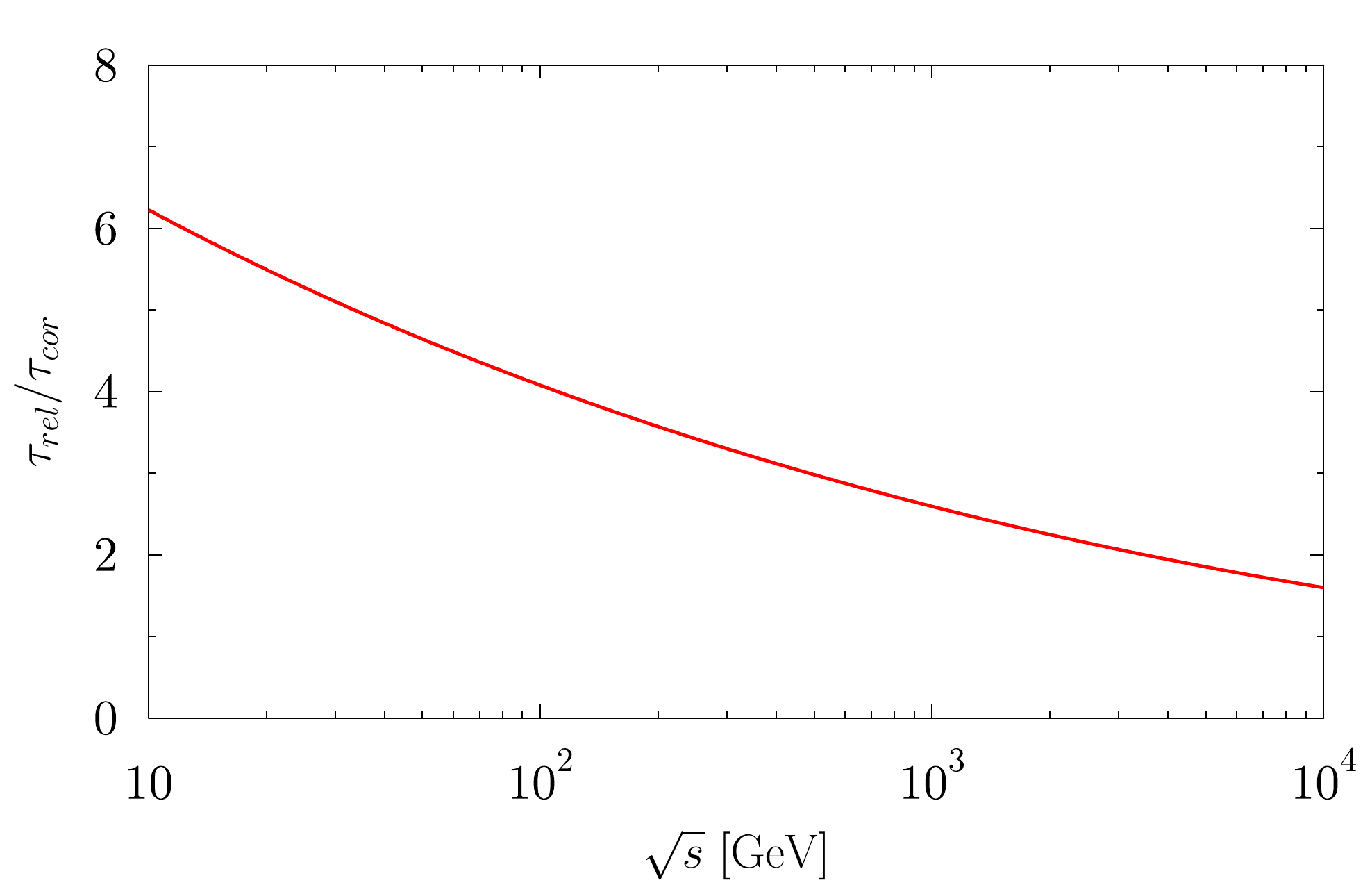}
\end{center}
\vspace{-5mm}
\caption{Energy dependence of $\tau_{rel}/\tau_{cor}$ obtained from experimental data. }
\label{F3}
\end{figure}
Using ${\rm Var}\left(N\right)$ and $\langle N\rangle$ values evaluated from the charged-particle multiplicity distributions for non-single-diffractive proton-proton (antiproton) collisions \cite{P(N)data,SWW} we obtain the ratio $t_f/\tau_{cor}$ shown in Fig. \ref{F2}. Combining the results of both approaches, we present in Fig. \ref{F3} the ratio $\tau_{rel}/\tau_{cor}$ in the energy range from $10$ GeV to $7$ TeV analyzed here.

\section{Interpretation of the results}
\label{Interpretation}

Let us now try to organize these results and draw some conclusions from them. The quasi-power distribution can be interpreted as a trace of temperature fluctuations \cite{WW} and the non-extensive Tsallis statistics, usually called superstatistics \cite{SST1,SST2}. If we approximate the production process with an irreducible Markov chain, then the dependence of fluctuations on time will be very sensitive to the reciprocal of the relaxation time, $\omega = 1/\tau_{rel}$, that is, the stochastic collision frequency for the particle \cite{HCA}. It is therefore reasonable to choose $\tau$ so that the fluctuation decay time along the particle's trajectory is the same as the decay time of a small section (small volume) of real matter surrounded by a much larger volume of its remnants. Now suppose that this small sample has a temperature variation such that its temperature is $T + \Delta T$. The sample will therefore gain or lose energy at a rate proportional to the temperature difference $\delta T$ and the thermal conductivity $\kappa$.

By dimensional analysis, it is easy to show that the rate of heat gain (in energy per time unit) is $\sim \kappa \Delta T$. Because each stochastic collision changes the system energy by amount $\sim \kappa \Delta T$ and the total stochastic collision frequency is $N\omega$, hence the rate of energy gain is $\propto N\omega \kappa \Delta T c_P$, where $c_P$ is the specific heat for constant pressure. By identifying the above increase in heat with the increase in energy, we obtain that the stochastic collision frequency for the particle (reciprocal of the relaxation time) is equal to \footnote{This can be compared to the result we get from the Fourier equation for heat transfer. $\frac{\partial T}{\partial t} = \frac{\kappa}{\rho c_P}\Delta T$, where $\rho$ is the density of the particles \cite{Landau}.}
\begin{equation}
\frac{1}{\tau_{rel}} = \omega = \frac{\kappa}{c_P}\frac{V}{N}. \label{omega}
\end{equation}
If the stochastic collisions are to simulate the effects of the surroundings of a set of $N$ particles, the frequency of the collisions should be as given in this formula. Note that $\omega$ is of the order $1/N$ and the total collision coefficient for the sample is of the order $1$. For a sufficiently large multiplicity $N$, the frequency of stochastic collisions will be much less than the frequency of inter-particle collisions. Therefore, most of the time most particles will move according to the conservative equations of motion for a closed system. The stochastic interruptions will be rare, but they will cause the system energy to relax to a value appropriate for temperature $T$ at a rate appropriate to the $N$ particle system, and will cause the energy to fluctuate around its equilibrium value with the magnitude appropriate for the canonical ensemble.

\begin{figure}[h]
\begin{center}
\includegraphics[scale=0.43]{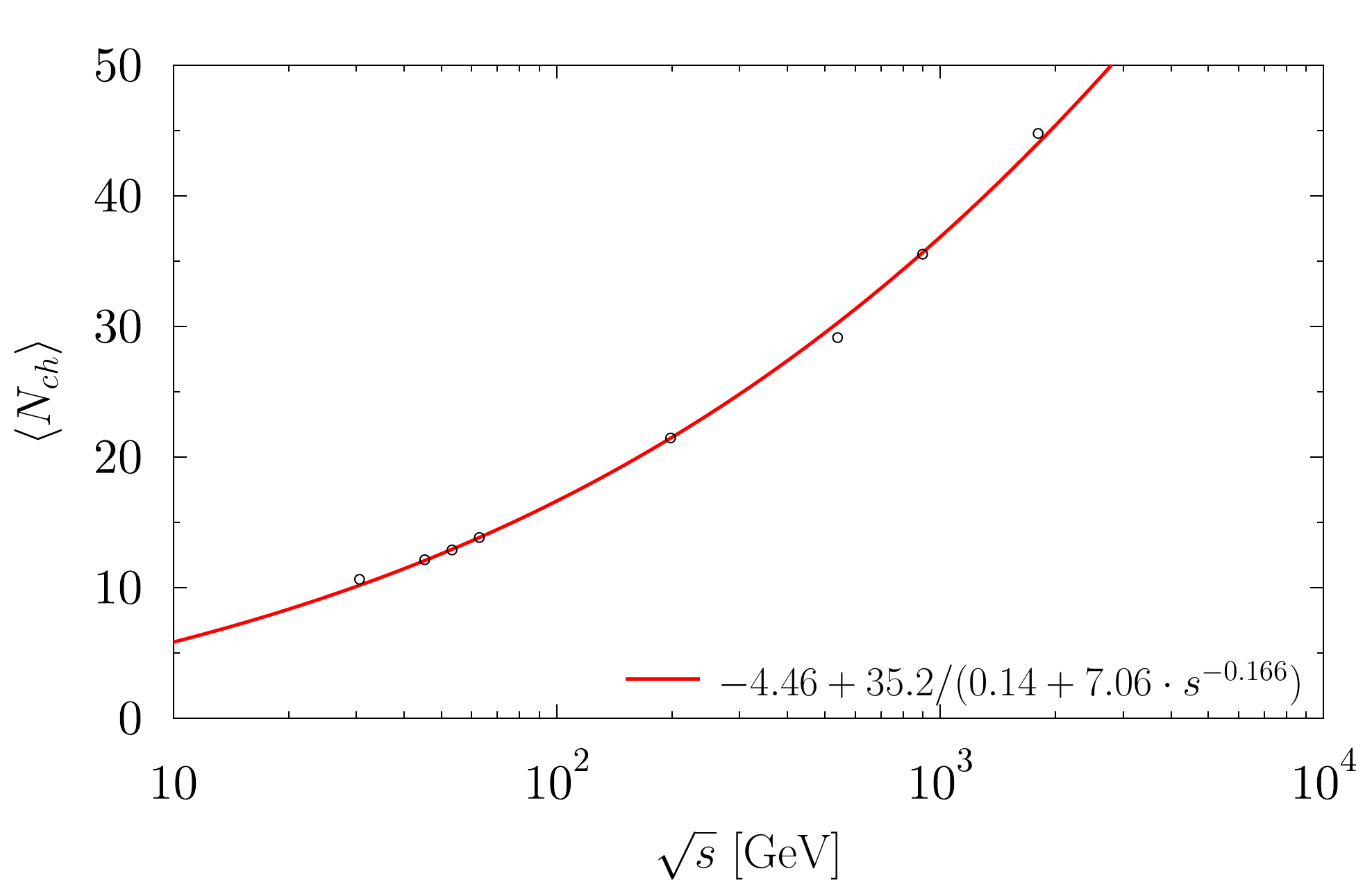}
\end{center}
\vspace{-5mm}
\caption{Energy dependence of mean multiplicity as given by Eq. (\ref{MeanN}). Data points are from \cite{P(N)data}.}
\label{F4}
\end{figure}

From Eq. (\ref{omega}) it can be expected that the multiplicity $N$ is related linearly to the relaxation time $\tau_{rel}$, $N \sim \tau_{rel}$. Assuming additionally that the freezout time $t_f$ is independent of energy, i.e. that the energy dependence shown in Fig. \ref{F1} comes only from the dependence of the relaxation time, $\tau_{rel} = \tau_{rel}(s)$ (which in turn comes from the energy dependence of the transverse momentum distributions), one can expect that for $\theta(s) = \frac{t_f}{\tau_{rel}}$ shown in Fig. \ref{F1} one gets
\begin{equation}
\langle N(s)\rangle = a + b/\theta(s). \label{MeanN}
\end{equation}
As can be seen in Fig. \ref{F4} comparison with data shows that this is indeed the case confirming the arguments presented above.
Note, however, that our result does not exclude the nonlinear dependence of $\tau_{rel}$ on $\langle N\rangle$. We show that $1/\theta \sim \langle N\rangle \sim (\sqrt{s})^b$. For $\tau_{rel} \sim \langle N\rangle^a$ we have energy dependent $t_f$, $t_f \sim (\sqrt{s})^{b(a-1)}$. For $a > 1$ $t_f$ increases with energy, while for $a<1$ it decreases with energy.

Interpretation of the results obtained from the analysis of multiplicity fluctuations is not so simple. The reason is that the fluctuations here come not only from the fluctuations of temperature $T$ but also from the fluctuations of the available energy $U$ (i.e. the fluctuations of inelasticity $K$). To ilustrate this we consider conditional Poisson distribution $P(N|n)$ with fluctuating mean value $\bar{n}$ according some distribution $w\left( \bar{n}\right)$. The resulting multiplicity distribution $P(N)$ is given by \footnote{Note that the exponential form of the distribution $w\left( \bar{n}\right)$ results in a geometrical (Bose-Einstein) distribution of $P(N)$, while for $w\left( \bar{n}\right)$ given by the gamma distribution we have the Negative Binomial Distribution of $P(N)$. }
\begin{equation}
P\left(N\right) = \int P\left( N|\bar{n}\right) w\left( \bar{n}\right) \bar{n} = \int \frac{\bar{n}^N}{N!} e^{-\bar{n}} w\left( \bar{n}\right) d\bar{n}. \label{PT}
\end{equation}
The fluctuations caused by $w\left( \bar{n}\right)$ define the moments of the distribution:
\begin{eqnarray}
\langle N\rangle &=& \langle \bar{n}\rangle, \label{x3}\\
{\rm Var}\left(N\right) &=& \langle N\rangle + {\rm Var}\left( \bar{n}\right), \label{x4}
\end{eqnarray}
and correspond to the correlation function by the relation
\begin{equation}
\langle \nu_2\rangle = \frac{{\rm Var}\left( \bar{n}\right)}{\langle \bar{n}\rangle^2}, \label{x5}
\end{equation}
The mean value of the distribution $w\left( \bar{n}\right)$ is
\begin{equation}
\langle \bar{n}\rangle = \frac{U}{T} = \frac{K \sqrt{s}}{T}, \label{x6}
\end{equation}
where both $T$ and $K$ can fluctuate, in fact we have that \cite{WW3}:
\begin{equation}
\frac{{\rm Var}\left( \bar{n}\right)}{\langle \bar{n}\rangle^2} = \frac{{\rm Var}\left(K\right)}{\langle K\rangle^2} + \frac{{\rm Var}\left(1/T\right)}{\langle 1/T\rangle^2}. \label{x7}
\end{equation}
Because ${\rm Var}\left(1/T\right)/\langle 1/T\rangle^{2} = q - 1$ (where $q$ is the nonextensivity parameter~\cite{WW3}) we have that
\begin{equation}
\frac{{\rm Var}\left(K\right)}{\langle K\rangle^2} = \langle \nu_2\rangle - (q-1). \label{x8}
\end{equation}
\begin{figure}[h]
\begin{center}
\includegraphics[scale=0.43]{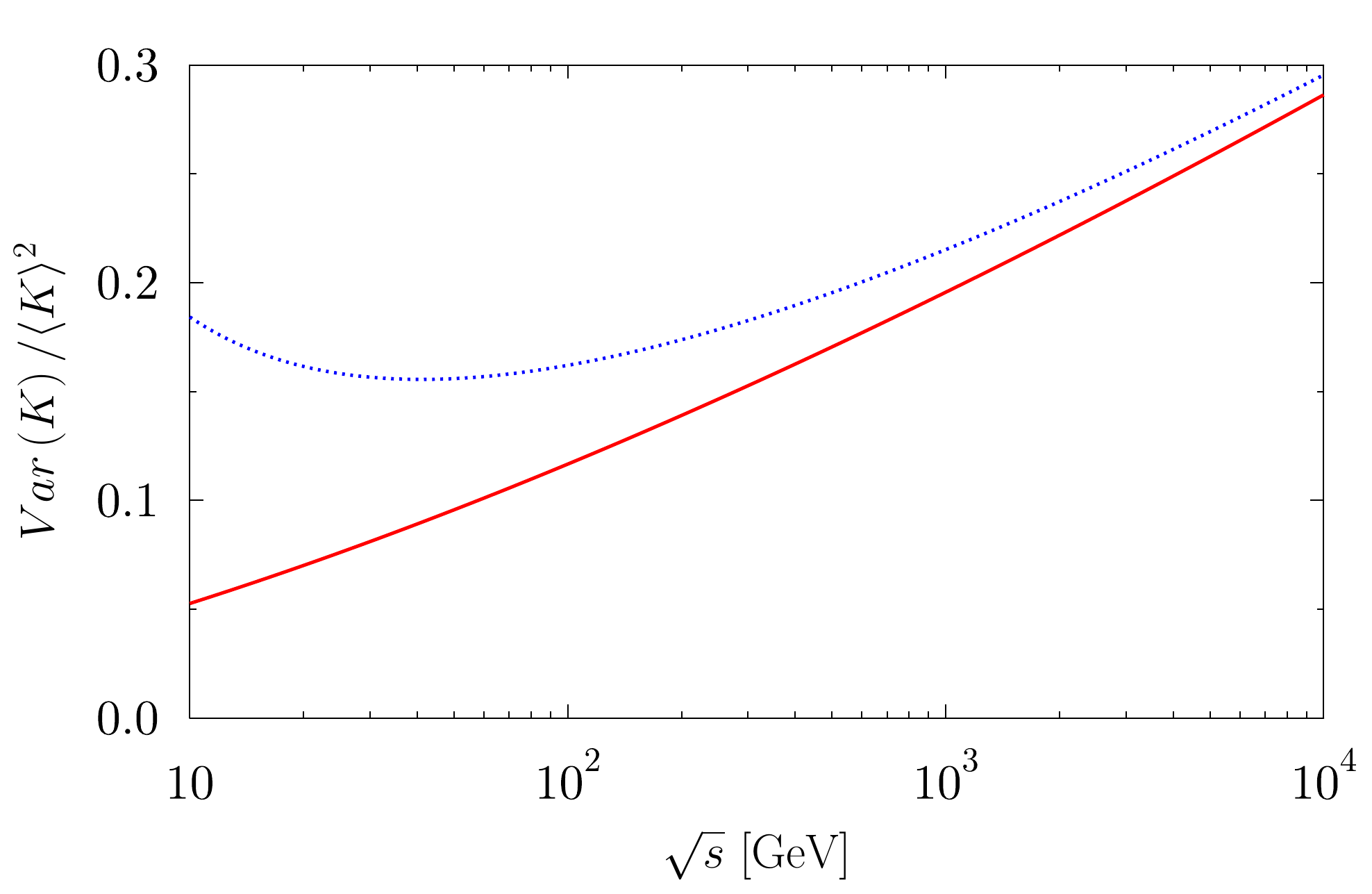}
\end{center}
\vspace{-5mm}
\caption{Relative fluctuations of the inelasticity coefficient emerging from the Eq. (\ref{x8}) (solid curve) and Eq. (\ref{x10}) with parameter $p=0.7$ (dotted curve).
}
\label{F6}
\end{figure}
Taking $q(s)$ used in Fig. \ref{F1} and $\langle \nu_2(s)\rangle$ such as in Fig. \ref{F2} we get the dependence on the energy of the relative fluctuations $K$ shown in Fig.~\ref{F6}. Because, approximately, $q-1 \cong \frac{1}{3}\langle \nu_2\rangle$ \cite{WW3}, therefore ${\rm Var}\left(K\right)/\langle K\rangle^{2}\simeq \frac{2}{3}\langle \nu_2\rangle$. For the uniform distribution of $K$, we have ${\rm Var}\left(K\right)/\langle K\rangle^2=1/3$ whereas for the symmetric triangle distribution ${\rm Var}\left(K\right)/\langle K\rangle^2=1/6$.  Inelasticity distribution at low energies $\sqrt{s}=16.5$ GeV show more or less triangle form~\cite{B1,B2} \footnote{The Feynmann $x$-spectrum of leading protons is close to a flat uniform distribution almost for all range, from $x=0$ to $x = 1$. The distribution of $K=1-\left(x_1+x_2\right))/2$ depends on the degree of correlation of the fractional energy contents $x_1$ and $x_2$ of the leading particles. If $x_1$ and $x_2$ are totally correlated, then $K$ is uniformly distributed, whereas if  $Cov\left(x_1,x_2\right)=0$ the distribution of $K$ is triangular.}.

However, the above results are for independent fluctuations of $K$ and $T$, and they can be correlated. In such a case the result can be quite different depending on the coefficient of correlations. Also, due to conservation rules, the initial distribution does not have to be Poisson distribution as in Eq. (\ref{PT}) but Binomial Distribution (BD). Then instead Eq. (\ref{x4}) we would have
\begin{equation}
{\rm Var}\left(N\right) = \langle N\rangle - \langle N\rangle p + {\rm Var}\left( \bar{n}\right) \label{x9}
\end{equation}
where $p$ is the emission probability which enters into the BD. In such a case we have
\begin{equation}
\frac{{\rm Var}\left(K\right)}{\langle K\rangle^2} = \langle \nu_2\rangle - (q-1) + \frac{p}{\langle N\rangle} \label{x10}
\end{equation}
which is shown in Fig. \ref{F6}.

\section{Summary and conclusions}
\label{Summmary}

To summarize, let us first note that from Figs. \ref{F1}, \ref{F2} and \ref{F3} one can deduce the relative positions of $t_f$, $\tau_{rel}$ and $\tau_{cor}$ relative to each other depending on the energy: for $\sqrt{s} < 570$ GeV $t_f > \tau_{rel} > \tau_{cor}$ whereas for $570$ GeV $< \sqrt{s} < 5.3$ TeV we have that $\tau_{rel} > t_f > \tau_{cor}$ and for $\sqrt{s} > 5.3$ TeV $\tau_{rel} > \tau_{cor} > t_f$. Relaxation time and correlation time are roughly related by the relation $\tau_{cor}\cdot t_f^{1/2} = 0.3 \left( \tau_{rel}\right)^{3/2}$.

The relaxation time $\tau_{rel}$ discussed in this paper does not describe the evolution of the distribution function $f$ as it is written in Eq. (\ref{SRBE}). In our case, it characterizes the time evolution of the non-extensivity parameter $q$ as shown in Eq. (\ref{A-3}). It therefore describes the temporal evolution of the temperature fluctuation, ${\rm Var}\left(1/T\right) = \frac{1}{4}\langle 1/T\rangle^2 \exp\left( - t/\tau_{rel}\right)$.
Nevertheless, presented by Eq. (\ref{sumrule}) method chosen in the Section \ref{Distribution} to determine the ratio $t_{rel}/\tau $, leads to the result given by Eq. (\ref{Result}), which is identical to what we get from Eq. (\ref{A-4}) resulting from Eq. (\ref{A-3}).

The dependence of $\tau_{rel}$ on energy mainly comes from the energy dependence of multiplicity (cf. Eq. (\ref{MeanN}) and Fig.~\ref{F4}). Note that the collision time $\tau_{coll}$, defined as $1/\tau_{coll} =\langle u \sigma\rangle N/V$, where $u$ is the thermal (relative) energy (relative) and $\sigma$ is the total cross-section for collisions between particles after averaging over the momentum, decreases with a multiplicity. Both of these times, $\tau_{rel}$ and $\tau_{coll}$, are related to each other by the relation $\tau_{coll}\tau_{rel} = c_P/(\langle u \sigma\rangle \kappa)$, which very weakly depends on energy. This suggests that collisions between particles play the role of a stochastic force causing temperature changes. And the temperature ($T$) fluctuations in combination with the inelasticity ($K$) fluctuations lead to multiplicity fluctuations. Thus, both of these fluctuations ($T$ and $K$) determine the correlation function $\langle \nu_2\rangle$  which is described by $\tau_{cor}$.

Finally, we note that in the fluctuating temperature scenario, the relaxation and correlation times are related to each other through the relationship of the transverse momentum distributions with the multiplicity distributions. As shown in~\cite{WW}, the fluctuation of inverse temperature given by the gamma distribution $w(1/T)$ leads to the replacement of the exponential distribution (\ref{BG}) by the Tsallis distribution of transverse momenta (\ref{TFit}) with the parameter $q = 1 + {\rm Var}\left(1/T\right)/\langle 1/T\rangle^{2}$. In the simplest case of a fixed available energy $U = const$, the fluctuation of $\bar{n} = U/T$ is given by the gamma distribution $w(\bar n)$ and leads, according to Eq. (\ref{PT}), to the NBD distribution for $P(N)$ described by the shape parameter $k$ such that  $1/k=q-1$ \cite{WW3}. In this case $\langle \nu_2\rangle = 1/k = q-1$. Moreover, in a single statistical ensemble, the relaxation time depends linearly on the mean multiplicity (cf. Eq. (\ref{omega}))
\footnote{ 
It is worth mentioning that if a similar relationship was also observed in the rare events associated with hard collisions, that have an multiplicity larger than minimum bias, this observation (corresponding to a
longer relaxation time, even in the case of a strong collision occurring earlier) has to do with the proposal that the hard scale of the collision is related to the thermal scale due to the entanglement of the proton wave function \cite{a,b}.}.
However, this does not apply to nuclear collisions. For example, in superposition models where secondary particles are emitted by independent $N_S$ sources generated by interacting nucleons we have that $N_{AA} = N_S N_{pp}$ and $\langle \nu_2\rangle _{AA} = \langle \nu_2\rangle_{pp} /\langle N_S\rangle + Var\left(N_S\right)/\langle N_S\rangle^2$. Many other scenarios are possible here as well (cf., for example, clustering processes~\cite{a,c}) leading to a different predictions.

\vspace*{0.3cm}
\centerline{\bf Acknowledgments}
\vspace*{0.3cm}
This research was supported in part by the National Science Centre (NCN) Grant 2016/23/B/ST2/00692 (MR) and by grants UMO-2016/22/M/ST2/00176 and DIR/WK/2016/2018/17-1 (GW).

\appendix*

\section{RTA and beyond}
\label{NonRta}

\begin{figure}[h]
\begin{center}
\includegraphics[scale=0.43]{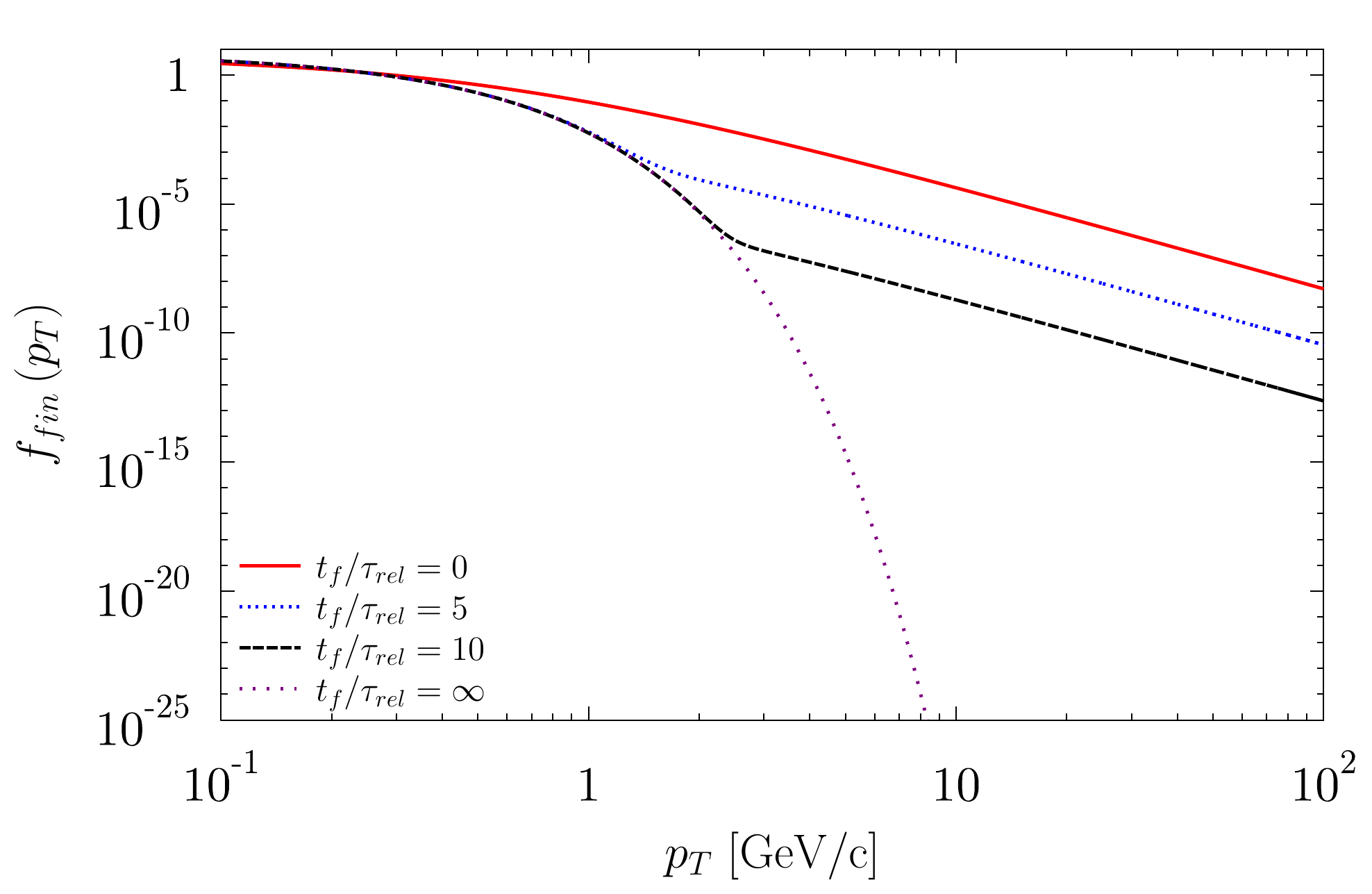}
\end{center}
\vspace{-5mm}
\caption{Schematic transverse momenta distributions $f_{fin}$ resulting from the relaxation time approximation scenario for
$q_{in}=1.25$, $q_{eq}=1.0$, $T=0.14$ GeV, and for $t/\tau = 0,~5,~ 10,~ \infty$ (curves from top to down).}
\label{FA1}
\end{figure}

Notice that Eq. (\ref{RTA-sol}) defining RTA can be rewritten as a two-component final distribution
\begin{equation}
f_{fin} = f_{in}\exp\left(-\frac{t_f}{\tau_{r el}}\right) + f_{eq}\left[ 1 - \exp\left(-\frac{t_f}{\tau_{rel}}\right)\right]. \label{A-1}
\end{equation}
Using $f_{in}$ and $f_{eq}$ in the form of Tsallis distribution (with, respectively, $q_{in}$ and $q_{eq}$) we get $f_{fin}$  for different values of $t_f/\tau_{rel}$ as presented in Fig. \ref{FA1}.
\begin{figure}[h]
\begin{center}
\includegraphics[scale=0.43]{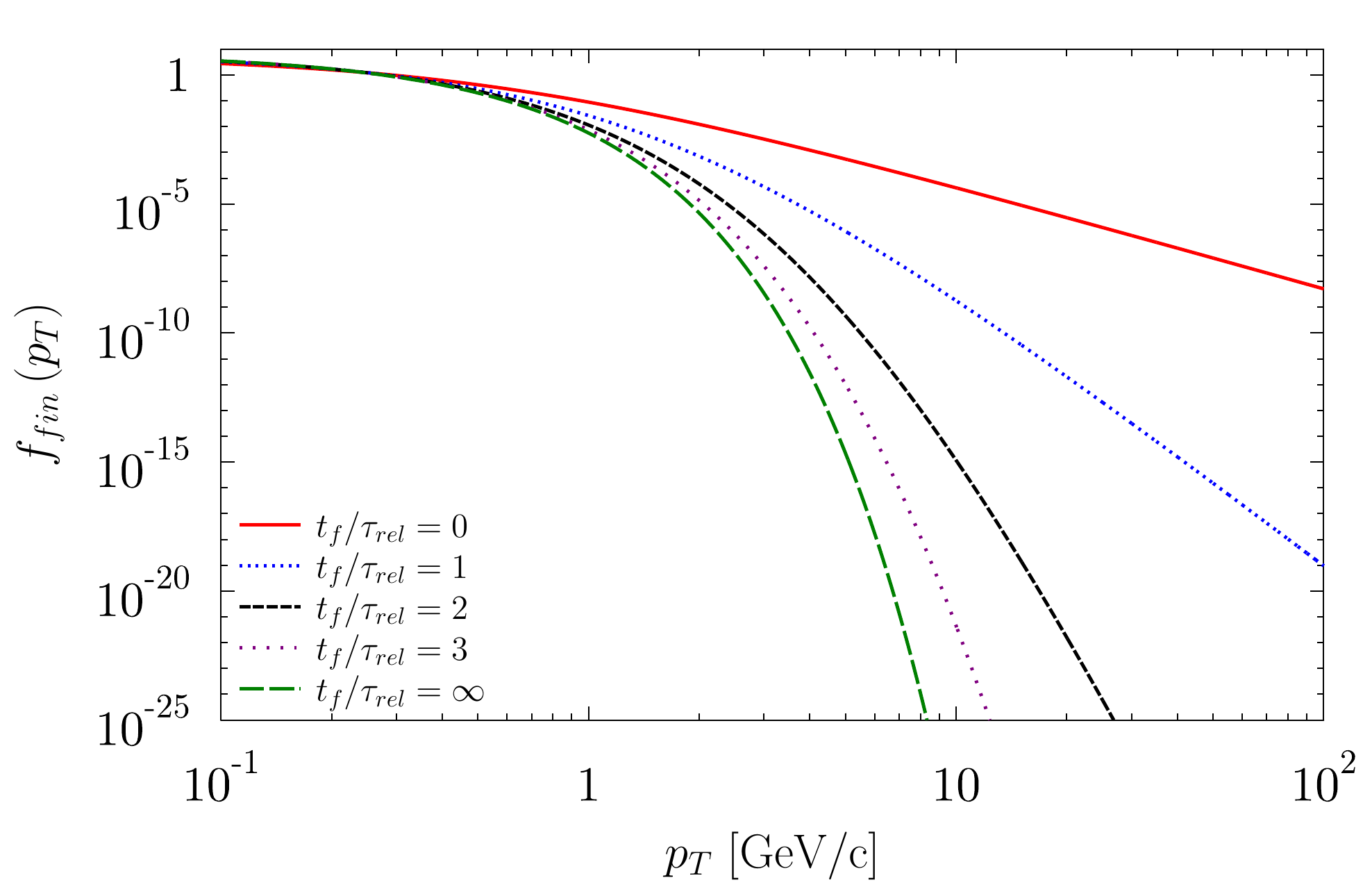}
\end{center}
\vspace{-5mm}
\caption{Schematic transverse momenta distributions for $f_{fin}$ calculated from Eq. (\ref{A-2}) for $t/\tau = 0,~ 1,~ 2,~ 3,~ \infty$ (curves from top to down). 
The values for $T$, $q_{in}$, $q_{eq}$ are the same as for Fig. \ref{FA1}. }
\label{FA2}
\end{figure}

However, if we would require that all distributions $f(t)$ in Eq.~(\ref{RTA-sol}) have the form of Tsallis distributions depending on time entirely via the time dependence of the corresponding nonextensivity parameters, $f(t) = f[q(t)]$, then the time evolution would be given by
\begin{equation}
\frac{\partial f(t)}{\partial t} = F[q(t)] \label{A-2}
\end{equation}
(with quite involved form of $F$ \footnote{The form of the function $F$ from Eq. (\ref{A-2}) can be deduced by taking $f(t)$ given by Tsallis distribution with $q = q(t)$ and calculating $df/dt$. As a result, we get that $F\left[p_T,q(t)\right] = f\left[ p_T,q(t)\right]\cdot\left\{ \ln\left[ 1 + (q(t)-1)\frac{p_T}{T}\right] + \frac{T}{T + [q(t)-1]p_T} - \frac{[q(t) - 1]^2 +1}{2 - q(t)} \right\}\cdot\frac{1}{[q(t)-1]^2}\frac{dq(t)}{dt}$. Note that approximately (because $\ln x + 1/x \approx 1$) we get that $\frac{\partial f}{\partial t} = \frac{-f}{(2-q)}\frac{d q(t)}{dt} =  \frac{f}{\tau_{rel}}\frac{q-1}{2-q}$.}). Assuming further that the dependence of $q$ on time is given by
\begin{equation}
\frac{\partial q}{\partial t} = \frac{q_{eq}-q}{\tau_{rel}}, \label{A-3}
\end{equation}
and remembering that we always assume that $q_{eq} =1$, we have that
\begin{equation}
q - 1 = \left( q_{in} - 1\right) \exp\left( - \frac{t_{f}}{\tau_{rel}}\right) \label{A-4}
\end{equation}
which corresponds to Eq. (\ref{Result}). Fig. \ref{FA2} shows the resultant schematic distributions $f_{fin}$ for different $t_{f}/\tau_{rel}$; they all have form of Tsallis distribution with $q=q\left(t=t_f\right)$ as given by Eq. (\ref{A-4}). As one can see the result now is different from that from the RTA approximation shown in Fig. \ref{FA1}.

\end{document}